\documentstyle[preprint,aps,epsfig]{revtex}
\draft

\begin{document}
%@twocolumnfalse\endcsname

\title{Surface Plasmon Dispersion Relations in Chains of Metallic
Nanoparticles: Exact Quasistatic Calculation}
\author{Sung Yong Park and David Stroud}
\address{Department of Physics,
The Ohio State University, Columbus, Ohio 43210}

\date{\today}

\maketitle

\begin{abstract}

We calculate the surface plasmon dispersion relations for
a periodic chain of spherical metallic nanoparticles 
in an isotropic host, including all multipole modes in a
generalized tight-binding approach.  
For sufficiently small particles ($kd \ll 1$, 
where $k$ is the wave vector and $d$ is the interparticle separation),
the calculation is exact. The lowest bands differ only 
slightly from previous point-dipole calculations
provided the particle radius $a \lesssim d/3$,
but differ substantially at smaller separation.
We also calculate the dispersion relations for many higher bands, 
and estimate the group velocity $v_g$ and the exponential decay length $\xi_D$
for energy propagation for the lowest two bands 
due to single-grain damping.  For $a/d=0.33$, the result for $\xi_D$ is in qualitative 
agreement with experiments on gold nanoparticle chains, while
for larger $a/d$, such as $a/d=0.45$, $v_g$ and $\xi_D$ are expected to be strongly
$k$-dependent because of the multipole corrections.  When $a/d \sim 1/2$,
we predict novel percolation effects in the spectrum,
and find surprising symmetry in the plasmon band structure.
Finally, we reformulate the band
structure equations for a Drude metal in the time domain, and suggest
how to include localized driving electric fields in the equations
of motion. 

\end{abstract}

%\pacs{PACS numbers: 78.67.Bf, 42.79.Gn, 71.45.Gm,73.20.Mf}
%\vskip1.5pc
%\newpage

\section{Introduction}

Recently, it has been shown that energy can be transmitted along 
a one-dimensional (1D) chain of equally spaced metallic nanoparticles, 
via propagating surface plasmon (SP) modes~\cite{quinten,krenn,brong,maier,maier1,maier2,maier3,gray}.  
These modes are basically linear combinations of single grain
SP modes, i.\ e. oscillations of electronic charge within a single 
grain\cite{kreibig}.
The single-grain modes are accompanied by an oscillating electric
moment (dipole and higher) on the grain.  The electric field of this moment in 
turn generates oscillating moments on neighboring spheres.

The propagating SP modes are simply traveling waves of
these oscillating  moments.  They are characterized by well defined
dispersion relations $\omega(k)$ which relate their frequencies 
$\omega$ and wave vectors $k$\cite{krenn,brong}.  If the damping is
sufficiently small, the energy transmitted by these ways may travel
at speeds up to 0.1c.  Thus, one can imagine a variety of possible uses
for these waves.

The calculation of $\omega(k)$ typically involves several approximations.
The first of these is the {\em near field}
approximation - that is, one assumes that $kd \ll 1$, where
$k$ is the wave number and $d$ the interparticle separation.  
This assumption permits the electric field ${\bf E}$ to be calculated in
the quasistatic limit, in which ${\bf E}$ is expressed as the gradient of a 
scalar potential.  

A second common approximation
is that the field produced by a given particle at its neighbors
is that of a {\em point dipole}.  However, this second assumption is stronger
than the quasistatic approximation, and becomes inaccurate when the
particles are closely spaced.  Under these conditions,  
the quasistatic fields may be modified significantly by multipolar 
interactions.  Typically, these multipolar fields
have been included by numerical techniques such as finite-difference 
time-domain calculations.  However, it may be difficult to obtain 
accurate results by these numerical techniques, because the higher multipole
excitations may produce fields which vary rapidly in space, whereas numerical
studies with insufficient discretization may not achieve adequate
resolution.  Thus, an exact calculation in a simple geometry may not only
provide physical insights into this system, but also give useful guidelines
for the validity of numerical calculations in more complex geometries.

In the present work, we will show how these multipolar corrections
can be calculated {\em exactly}, using a straightforward analytical
approach.  The formalism is analogous to the generalized tight-binding
method in conventional band theory.  In this approach, one constructs
Bloch states from individual atomic orbitals, and then diagonalizes the
Hamiltonian matrix in the basis of these atomic orbitals.  In the surface
plasmon analog, the individual atomic orbitals are multipolar SP oscillations
for each sphere.  The matrix elements needed to calculate the Hamiltonian
matrix are easily constructed, especially for a periodic one-dimensional (1D)
chain of spheres.  The diagonalization needed to calculate the bands
is readily carried out.  The entire calculation is made simpler in
1D systems, because the Hamiltonian matrix decomposes into separate blocks,
one for each azimuthal quantum number $m$.         

The basic formalism necessary to carry out this calculation\cite{bergman}  
has thus far been applied only rather sparingly, because there have been few
experimentally available realizations of the ordered geometry required for
this approach.  It has been applied mainly to calculating
the effective complex dielectric function $\epsilon_e(\omega)$ of
a periodic composite medium, which requires the 
SP band structure only at Bloch vector ${\bf k} = 0$.  
Only recently has it become possible to produce well-controlled ordered
metallic structures at the nanoscale, and hence to generate and detect
these SP waves at general wave vectors.  In this paper, we describe and
numerically solve the equation necessary to calculate this band 
structure in the general case of ${\bf k} \neq 0$ in 1D systems.

Besides giving the solutions in the full multipolar case, we also include
damping of the SP modes due to losses within the individual metallic
particles.  For a Drude metal, these losses can be treated by including
a finite relaxation time $\tau$ in the Drude dielectric function.  
This damping can be included approximately by 
adding an imaginary part to the SP frequencies.  We also briefly
discuss why the damping due to radiative energy losses is expected to be small.
These losses arise from the breakdown of the quasistatic approximation,
and can, in principle, also be approximately included by adding an imaginary 
part to the surface plasmon frequency.

We will also present the multipolar SP equations in the {\em time} domain for
a Drude metal, where they take the form of a set of coupled second order
ordinary differential equations.   In this form, it is straightforward to 
include single-particle damping (and also, in principle, radiative damping).
Moreover, one can also incorporate driving terms, arising, e.\ g., from
external electric fields.  These equations may thus be useful in modeling
specific types of experimental probes which produce localized time-dependent
electric fields.    

The remainder of this paper is organized as follows.  In Section II, we
describe the formalism needed to calculate the SP band structure, and
specialize to the calculation for a 1D chain.  Section III presents numerical results 
in this 1D chain system.   
In Section IV, we discuss our results, and suggest some possible extensions to other 
geometries.
Finally, the Appendix presents an alternative formulation of the equations of motion 
in the time domain, and describes how localized time-dependent driving electric fields 
can be included in these equations.   

\section{Formalism}

We consider a 1D chain of spherical particles of radius $a$, separated
by a distance $d$ ($d \geq 2a$).  The particles are assumed to be
arranged along the $z$ axis with centers at $z = 0$, $\pm d$, 
$\pm 2d$, $\cdots$.   We assume that the particles and host have 
dielectric functions
$\epsilon_m(\omega)$ and $\epsilon_h(\omega)$.  
To be definite, we may consider the particles as
metallic and the host as insulating, but the discussion below
applies to any choice of $\epsilon_m$ and $\epsilon_h$.  
All the formalism is given in terms of a frequency variable $s$ defined by
\begin{equation}
s = \frac{1}{1 - \epsilon_m/\epsilon_h}.
\label{eq:s}
\end{equation}
As will be seen below, and as is discussed indirectly in Ref.\ \cite{bergman},
all the allowed propagating SP frequencies correspond to $s$ in the
range $0 \leq s \leq 1$, or equivalently, to $-\infty \leq
\epsilon_m/\epsilon_h \leq 0$,  assuming that $\epsilon_m$ and $\epsilon_h$ are both real.  

We will calculate the SP band structure in an ``atomic'' basis $n, \ell, m$.  
Here, $\ell$ and $m$ label the ``angular momentum'' of the eigenfunction, 
and $n$ labels the grain.
Thus the allowed values of these indices are 
$n = 0, \pm 1, \pm 2, \cdots$, $\ell = 1, 2, 3, \cdots$, and $m = -\ell, -\ell + 1, \cdots , \ell$.
In order to calculate the SP band structure in this basis, we first need the
matrix element $Q_{n\ell m; n^\prime \ell^\prime m^\prime}$, where 
$n \neq n^\prime$.  This matrix element is given by (see, e.\ g., Ref.\ 
\cite{bergman})
\begin{eqnarray}
Q_{n\ell m; n^\prime\ell^\prime m^\prime} = 
(-1)^{\ell^\prime+m^\prime}
\left(\frac{a}{|n - n^\prime|d}\right)^{\ell+\ell^\prime + 1}
\left(\frac{\ell\ell^\prime}{(2\ell+1)(2\ell^\prime+1)}\right)^{1/2}
\nonumber \\
\times
 \frac{(\ell+\ell^\prime+m-m^\prime)!}{[(\ell+m)!(\ell^\prime+m^\prime)!
(\ell-m)!(\ell^\prime -m^\prime)!]^{1/2}}.
\exp(i\phi_b(m^\prime-m))P_{\ell^\prime+\ell}^{m^\prime-m}(\cos\theta_b).
\end{eqnarray}
Here we have introduced ${\bf b} = (n^\prime - n)d\hat{z}$, which
is the vector separation between the grains centered at $n^\prime\hat{z}$ and
$n\hat{z}$, and the polar and azimuthal angles
$\theta_b$ and $\phi_b$ for this displacement vector.
Since ${\bf b} = b\hat{z}$, $\theta_b$ is either $0$ or $\pi$, 
depending on whether $b$ is positive or negative.  If $b > 0$, 
$P_{\ell^\prime+\ell}^{m^\prime-m} = \delta_{m^\prime,m}$, whereas if
$b < 0$, $P_{\ell^\prime + \ell}^{m^\prime-m} = (-1)^{\ell^\prime+\ell}
\delta_{m^\prime,m}$.  Incorporating these simplifications, we
find that with
\begin{eqnarray}
Q_{n \ell m; n^\prime \ell^\prime m^\prime} =
(-1)^{\ell^\prime + m}\left(\frac{a}{|n^\prime-n|d}\right)^{\ell+\ell^\prime+1}
\left(\frac{\ell\ell^\prime}{(2\ell+1)(2\ell^\prime+1)}\right)^{1/2}
\nonumber \\
\times
\frac{(\ell+\ell^\prime)!}{[(\ell+m)!(\ell^\prime+m)!
(\ell-m)!(\ell^\prime-m)!]^{1/2}}\left(\frac{n^\prime-n}{|n^\prime - n|}
\right)^{\ell+\ell^\prime}\delta_{m,m^\prime}.
\label{eq:qlmn}
\end{eqnarray}
The matrix elements are diagonal in $m$ because the one-dimensional chain
is unchanged on rotation by any angle about the $z$ axis.

Next, we define the matrix element
\begin{equation}
Q_{\ell m; \ell^\prime m^\prime}(k) =
\sum_{n \neq 0} Q_{0,\ell m; n\ell^\prime m}\exp(inkd)\delta_{m,m^\prime},
\label{eq:qk}
\end{equation}
where the sum runs over all positive and negative integers except 
$n = 0$.  We have used the fact that, for this periodic 1D system, 
$Q_{n\ell m;n^\prime \ell^\prime m^\prime}$ is a function only of
$n - n^\prime$ and vanishes for $m \neq m^\prime$.  
After a little algebra, using Eqs.\ (\ref{eq:qlmn}) and (\ref{eq:qk}), 
we obtain
\begin{equation}
Q_{\ell m; \ell^\prime m^\prime}(k) = \left(\frac{a}{d}\right)^{\ell+\ell^\prime+1}
\sum_{n = 1}^\infty\frac{\cos(nkd)}{n^{\ell+\ell^\prime+1}}
K_{\ell,\ell^\prime,m}\delta_{m,m^\prime}
\end{equation}
for $\ell + \ell^\prime$ even, and
\begin{equation}
Q_{\ell m; \ell^\prime m}(k) = \left(\frac{a}{d}\right)^{\ell+\ell^\prime+1}
\sum_{n = 1}^\infty\frac{\sin(nkd)}{n^{\ell+\ell^\prime+1}}
K_{\ell,\ell^\prime,m}
\delta_{m,m^\prime}
\end{equation}
for $\ell+\ell^\prime$ odd, where
\begin{equation}
K_{\ell,\ell^\prime,m} = 2(-1)^{\ell^\prime+m}\left(\frac{\ell\ell^\prime}
{(2\ell+1)(2\ell^\prime+1)}\right)^{1/2}
\frac{(\ell+\ell^\prime)!}{[(\ell+m)!(\ell^\prime + m)!
(\ell-m)!(\ell^\prime-m)!]^{1/2}}.
\end{equation}

Finally, in terms of the matrix elements $Q_{\ell m \ell^\prime m}$; 
the SP structure is given by
\begin{equation}
det|s - H| = 0,
\label{eq:det}
\end{equation}
where the ``Hamiltonian'' $H$ has matrix elements
\begin{equation}
H_{\ell m;\ell^\prime m}(k) = s_\ell\delta_{\ell,\ell^\prime}
+Q_{\ell m; \ell^\prime m}(k), 
\label{eq:matrix}
\end{equation}
and the ``atomic'' eigenvalues $s_\ell$ are given by
\begin{equation}
s_\ell = \frac{\ell}{2\ell+1}.
\end{equation}
Note that $s_1 = 1/3$, while $s_\ell \rightarrow 1/2$ as $\ell \rightarrow
\infty$.

The full SP band structure at wave vector $k$ 
is obtained by diagonalizing the matrix (\ref{eq:matrix}).  As in conventional band
theory, there are many band energies for a given $k$, and one need
consider only the bands in the first
Brillouin zone, i.\ e., in this case, for $-\pi/d < k \leq \pi/d$, since the
states at other values of k are equivalent to those in the first zone.
For this one-dimensional system, the Hamiltonian breaks into
separate blocks, one for each value of $m$; this conveniently reduces the
size of the matrix which needs to be diagonalized.  Finally, as in the
LCAO (linear combination of atomic orbital) method of conventional band
theory, the band structure that results from this analysis is composed
of bands which originate from various ``atomic'' orbitals.  In the present
case, the the ``atomic'' states are multipolar SP modes associated with
the individual spheres.  These are degenerate at different $m$ (since the
individual particles are spheres), and have eigenvalues 
$s_\ell = \ell/(2\ell+1)$.

The band structure that results from diagonalizing the matrix (\ref{eq:matrix}) 
is expressed in terms of the variable $s$.  Thus, the bands have the
form $s_\alpha(k)$, where $\alpha$ labels the individual bands.  These
may be converted into frequencies using the relation (\ref{eq:s}).  
This dispersion relation will take on various forms, according to how $\epsilon_m$ 
(and $\epsilon_h$) depend on $\omega$.   
Here we assume that the system consists of Drude metal particles in vacuum, so that
$\epsilon_h = 1$ and $\epsilon_m(\omega) = 1 - \omega_p^2/[\omega(\omega+i/\tau)]$.
If $\omega_p\tau \rightarrow \infty$, then the appropriate 
conversion is given by 
\begin{equation}
\omega_\alpha(k) = \omega_p\sqrt{s_\alpha(k)}.
\label{eq:drude}
\end{equation}
For $\ell = 1$ there are three degenerate modes at frequency
$\omega_p/\sqrt{3}$, while for $\ell \rightarrow \infty$,
the modes approach the limiting value of $\omega_p/\sqrt{2}$.

If the relaxation time is finite, then the relation
\begin{equation}
s = \frac{1}{1 - \epsilon_m/\epsilon_h} = 
\frac{\omega(\omega + i/\tau)}{\omega_p^2}
\label{eq:drude1}
\end{equation}
can be inverted to express $\omega$ as a function of $s$.  For a real $s$,
the corresponding $\omega$ has both a real and imaginary part.  Thus,
we can map an eigenvalue $s_\alpha(k)$ (where $\alpha$ is a band index)
to a {\em complex} eigenvalue
\begin{equation}
\omega_\alpha(k) = \sqrt{\omega_p^2s_\alpha(k) - 1/(4\tau^2)}-i/(2\tau).
\label{eq:solution}
\end{equation}
The imaginary part describes the damping of this mode due to the finite
lifetime of the surface plasmons in the individual spheres.  If 
$\omega_p\tau \gg 1$, this damping is small, and the shift due to the
damping (as embodied in the factor $1/\tau^2$ within the square root)
is even smaller.  
Note that we have not included radiative damping in this expression.
In contrast to single-particle damping, the radiative damping depends on
the particle size, being greater for larger particles.  For 10-nm radius
gold spheres, it has been estimated that
only 1.5\% of the total damping rate is due to radiative
damping~\cite{sonnichsen}.  Also, according to Refs.\ \cite{brong} 
and \cite{maier}, the radiative damping should be small for particles of 
such a size because of strong near-field interactions. 

In the Appendix, we present an 
alternative formulation of the equations of motion in the time domain to obtain the SP
band structure, assuming a Drude dielectric function. In this formulation, which has the 
advantage that it can deal with localized time-dependent driving electric fields, these
radiative corrections could easily be included, as has been discussed, for example, in
Ref.~\cite{brong}.

In order to compare with experiment, we will consider two more quantities which can be
obtained from the dispersion relation $\omega(k)$: First, the $k$-dependent group velocity 
$v_g(k)$ is given by the relation
\begin{equation}
v_g(k) \equiv \frac{d\omega}{dk},
\label{eq:vg}
\end{equation} 
and can be easily computed numerically, given $\omega(k)$.
Secondly, we can also use Eqs.\ (\ref{eq:solution}) and (\ref{eq:vg}) to
estimate the energy loss from a plasmon propagating along a chain, which
is important in applications.   For this purpose, we 
define energy decay lengths, $\xi_D(k)$ for the lowest longitudinal and 
transverse modes, as the distance over which the energy density in the wave 
amplitude decreases by a factor $\exp (- 1)$.
If the complex band frequency is denoted $\omega_1(k) + i\omega_2(k)$, 
then $\xi_D(k)$ is defined by
\begin{equation}
\frac{\xi_D(k)}{d} = \frac{v_g(k)}{2\omega_2 d} = \frac{1}{2 \omega_2}
\frac{d\omega(k)}{d(kd)}.
\label{eq:xi}
\end{equation}
Note that in the case of the Drude approximation, the imaginary part of the complex
band frequency does not depend on $k$, and thus $\xi_D$ is just
proportional to $v_g(k)$. 

\section{Numerical Results}

We have diagonalized the matrix (\ref{eq:matrix}) to obtain the surface plasmon 
band structure for various values of the parameter $a/d$.   We include
all bands up to  $\ell = 80$, which is sufficient to insure convergence
of $s_\alpha(k)$ to within 1\%. 
The results are shown in Fig.\ 1.  
For comparison, we also show the results of truncating the Hamiltonian
matrix at $\ell = 1$.   
In this latter case, the Hamiltonian matrix is a diagonal $3 \times 3$
matrix with elements
\begin{equation}
H_{1m;1m}(k) = \frac{1}{3} + \left(\frac{a}{d}\right)^3\sum_{n = 1}^\infty
\frac{\cos(nkd)}{n^3}K_{1,1,m},
\end{equation}
where $K_{1,1,0} = -4/3$ and 
$K_{1,1,\pm 1} = 2/3$.
The corresponding plasmon bands, expressed as $s_{\alpha}(k)$, are shown
as open square ($m = 0$) and circle ($m= \pm 1$) in Fig.\ 1.  
If we use the Drude expressions $\omega_\alpha^2(k) = \omega_p^2s_\alpha(k)$, 
then these correspond exactly to the dipolar SP band structures 
obtained in Ref.~\cite{brong}.  This behavior is as expected, since when we 
retain only the $\ell = 1$ terms in the Hamiltonian, we are neglecting all the 
quasistatic contributions except the dipole fields.

As is evident, the plasmon bands take on quite a different form when
the higher values of $\ell$ are included in the Hamiltonian matrix.
For small values of $a/d$ (i.\ e. $a/d \lesssim 0.3$),
the lowest three bands are very similar to the bands obtained when
only the $\ell = 1$ matrix elements are included.  
This behavior is not surprising, since at these
values of $a/d$ the matrix elements connecting the $\ell = 1$ states to
those of higher $\ell$ are quite small, for any value of $k$.  This
smallness originates in the factors of $(a/d)^{\ell+\ell^\prime + 1}$
which appear in all the expressions for the matrix elements.  The smallest
factor connecting $\ell = 1$ to higher $\ell $ is $(a/d)^4$ for $k \neq 0$,
and $(a/d)^5$ for $k = 0$.  Thus, for relatively small values of $a/d$,
these connecting matrix elements are substantially smaller than the diagonal
ones.  

When $a/d \gtrsim 0.35$, the interband matrix elements start to become
substantial.  When this happens, the shapes of the lowest bands start to depart
significantly from the purely dipolar form seen for smaller $a/d$.
As is evident, by the time
$a/d \rightarrow 1/2$, the band structures of the lowest bands are so
altered that they no longer bear any obvious relation to these dipolar band
shapes.  Precisely at $a/d = 1/2$, the lowest state
at $k = 0$ reaches the limiting value $s = 0$.  This behavior
is a percolation effect: when $a/d = 1/2$, the two neighboring
spheres just touch, and the entire line of spheres becomes one connected
chain.  Thus, one might expect that the lowest eigenvalues of this chain
would resemble that of a very long cylinder, which indeed has as its lowest
eigenvalue $s = 0$.   

The band structure also acquires a striking symmetry near
$a/d  = 1/2$.  First, there appears to be a nearly perfect
reflection symmetry about the line $s = 1/2$.  In
addition, there is another reflection symmetry about the 
line $k = \pi/(2d)$, i.\ e. at the middle value of $k$ in the 
first Brillouin zone.  As particular
examples of these symmetries, there appear to be 
eigenvalues of $s = 0$ and $s = 1$ at $k = \pi/d$, 
just as there are at $k = 0$.
We do not fully understand the reasons for these symmetries.  
The  $s \sim 0$ eigenvalue at $k = \pi/d$
apparently corresponds to a longitudinal mode (dipole moment of the spheres
parallel to the $z$ axis) in which each sphere oscillates 180$^o$ out of phase
with its neighbors.  The multitude of modes near $s = 1/2$ 
presumably
originate in the high-$\ell$ ``atomic'' modes, 
which have eigenvalues approaching $s = 1/2$.   

In Fig.\ 2, we show the eigenvalues of the two lowest bands at $k = 0$ 
plotted as a function of $a/d$. Here, the lowest band corresponds to 
longitudinal mode ($m = 0$) and the second lowest band to degenerate 
transverse modes ($m = \pm 1$). We performed two different calculations: 
In the first calculation, shown as open circles and squares, 
we assumed the dipole approximation and included only the $\ell = 1$ part of the 
Hamiltonian matrix. This calculation corresponds to the tight-binding approximation used 
in Ref.\ \cite{brong}.  In the second, we included all bands up to $\ell = 80$, 
which is sufficient to ensure the convergence of these two bands, as
in Fig.\ 1, and this inclusion of the higher multipoles 
starts to make a substantial difference for $a/d \gtrsim 0.35$.

The inset to Fig.\ 2 shows the splitting $\Delta s$ between the 
longitudinal and transverse modes at $k = 0$, plotted as a function of $a/d$.  
In the dipole approximation (dashed line), this splitting increases 
monotonically as $a/d$ increases.  However, as shown by the solid line 
in the inset, when the higher multipoles are included, the splitting 
reaches a maximum near $a/d = 0.46$, then decreases again.  

In order to compare with experiment, one needs to re-express the band structure as 
dispersion relations for $\omega(k)$ rather than $s(k)$, using Eq.~(\ref{eq:s}).  
With the resulting $\omega(k)$, we can also obtain $v_g(k)$ from Eq.~(\ref{eq:vg}) and
$\xi_D$ from Eq.~(\ref{eq:xi}).
We show the resulting dispersion relations $\omega(k)/\omega_p$ in Fig.\ 3 (a),
and the resulting $\xi_D(k)$ and $v_g(k)$ in Fig.\ 3 (b)
for the lowest longitudinal and transverse bands as a function of $kd$.  
We denote these results as open square and circle for $a/d = 0.33$, 
the value used in experiments, and also as solid and dashed lines 
for $a/d=0.45$, which is near the maximum of the splitting $\Delta s$.  
In order to calculate $\omega(k)/\omega_p$, we choose $\omega_p=6.79\times 
10^{15}$ rad/s and $\tau = 4$ fs, as used in Ref.~\cite{maier3}. 
This choice allows us to compare the present result for $a/d=0.33$ 
with those given in Refs.~\cite{maier1} and \cite{maier3}. 

First, we compare our results for $a/d=0.33$ with experiment.
For $a/d=0.33$, the result of the full calculation for $\omega(k)$ is not significantly 
different from the dipole approximation, since multipole effects produce only 
a minor alteration to the lowest bands in this case. 
However, the multipole effects can be seen much more clearly in the $k$-dependent
group velocity $v_g(k)$ for these bands, and this quantity can be easily computed 
numerically, using the $\omega(k)$ shown in Fig.\ 3 (a).  
In contrast to the result from the dipole approximation of
Ref.\ \cite{brong}, the maximum in $v_g$ for $a/d=0.33$ does not occur  
at $k=\pi/(2d)$, but instead around $k=\pi/(4d)$, when the multipolar
corrections are included.  However, if we assume $d=75$ nm, 
$\omega_p=6.79\times 10^{15}$ rad/s, and $\tau = 4$ fs as in 
Ref.~\cite{maier3}, the {\em magnitude} of the maximum $v_g$ for the
longitudinal ($m = 0$) mode is approximately $1.9 \times 10^7$ m/s, which is close to 
the result of Ref.\ \cite{maier3},
while the magnitude of $v_g$ for the transverse ($m=\pm 1$) modes is slightly 
larger than the value ($1.1 \times 10^7$ m/s) estimated in Ref.\ \cite{maier3}.
Also, the values of $\xi_D$ in the lowest longitudinal and transverse modes for 
$a/d = 0.33$ are comparable to the experimental values for gold, 
as given in Ref.~\cite{maier2}.

For $a/d=0.45$, which is near the maximum of the splitting $\Delta s$,  
the multipole corrections to the band structure are much greater.
As $a/d$ approaches the maximum splitting between longitudinal ($m = 0$) 
and transverse ($m = \pm 1$) modes, the variation of $v_g$ with $k$ becomes non-monotonic.
In contrast to the dipole approximation, which gives a maximum group 
velocity at $k=\pi/(2d)$, our exact calculation actually gives
a local {\em minimum} in $v_g$ for this value of $k$ (for both polarizations). 
As can be seen from Fig.\ 3(b), $v_g(k)$ has, in fact, two maxima as a function
of $k$ for this separation, for both longitudinal and transverse modes.  
The maximum estimated exponential decay length shown in Fig.\ 3(b), 
for the optimum $k$, corresponds to an $m = 0$ wave, and is about 
three times larger than that for $a/d=0.33$.  But this decay length is
calculated for a wave with $k$ vector corresponding to the maximum
group velocity.  The actual $v_g$ is strongly $k$-dependent, 
especially for the larger $a/d$.
Thus a typical wave, which would likely propagate as a packet of many 
different wave vectors, would likely have a quite different decay length,
and also would probably not decay exponentially.   
It is possible that this $k$-dependence is related to the non-exponential
spatial decay of SP's found in the numerical simulations of Ref.\
\cite{quinten}.

We have not commented thus far about the role of the higher SP bands.
For values of $a/d$ greater than about 0.33, most of these
bands are nearly dispersionless, with eigenvalues $s_\alpha(k) \sim 1/2$.  
The SP modes corresponding to these bands will thus propagate with very small 
group velocity $v_{g,\alpha} = d\omega_\alpha(k)/dk$, and are likely to
contribute very little to energy transport along the chain.

\section{Discussion}

In this work, we have calculated the surface plasmon dispersion relations
for a one-dimensional chain of spherical particles in a uniform host.
In contrast to previous work, we have included all the multipolar terms
in our calculation, within the quasistatic approximation.
We find that the dipole approximation is reasonably accurate 
for $a/d \leq 1/3$, but becomes increasingly inaccurate
for larger $a/d$.  When $a/d \rightarrow 1/2$,
the lowest band shapes are entirely different from
the point-dipole predictions.  Thus, an accurate comparison of theory to
experiment should take into account these corrections, when $a/d$ exceeds
about 1/3.

In our results near $a/d = 1/2$ we see conspicuous percolation effects.
Specifically, the $k = 0$ mode 
approaches zero frequency for a chain of 
Drude spheres in an insulating host.  
Furthermore, when $a/d \rightarrow 1/2$, the entire band structure
shows remarkable reflection symmetry, both around $k = \pi/(2d)$
and around the frequency midpoint at $s = 1/2$.  We do not presently
understand the reasons for this symmetry.

Besides producing shape distortions in the lowest bands, the present
calculations also lead to an infinite number of higher propagating
SP bands.  We believe, however, that these will contribute little
to energy propagation, because they are characterized by much lower
group velocities than the lowest two bands.

Our calculations in Fig.\ 3 have, of course, been carried out
in the Drude approximation.  As mentioned earlier, we used values of
$1/\tau$ and $\omega_p$ as best fits to experiments on bulk gold,
as described in Ref.\ \cite{maier3}.  In actuality, the complex dielectric
functions of silver, and especially gold, have substantial interband 
contributions, and cannot be described by a Drude dielectric function in the
visible.  
An accurate translation of the SP band structure from the variable $s$
to the variable $\epsilon_m(\omega)/\epsilon_h$ should use this more
accurate dielectric function, e.\ g. by using a fit of the experimental
$\epsilon_m(\omega)$ to a sum of free-electron and Lorentz oscillator parts.
This more complicated procedure might somewhat change the plasmon 
band structures, especially for gold.
Another possible complication is that, in typical experiments, the
nanoparticle chains are laid down on a substrate, whose dielectric constant
differs from that of vacuum.  Thus, the chain is not really embedded in a
homogeneous dielectric.  Some workers have taken this complication into
account by treating the host as homogeneous but with a dielectric function
which is an average over the air and substrate dielectric functions~\cite{maier1}.
Once again, this correction, if included, would also modify the calculated 
SP band structure.

The present work could be readily be generalized to higher dimensions,
e.\ g. to an ordered layer of spheres deposited on a substrate.   This extension
should be straightforward, since the matrix elements required are
the same as used here.  The same approach could also be applied to 
particles of different shapes, e.\ g. ellipsoids or 
short cylinders, although the calculation of the single particle 
eigenstates and the overlap integrals might be more difficult.
We plan to carry out some of these extensions in the near future.

\section*{Acknowledgments}

This work was supported by NSF Grant No. DMR01-04987, and 
benefited also from the computational 
facilities of the Ohio Supercomputer Center.

\section*{Appendix: Formulation in the Time Domain}

If the small spherical particles are described by a Drude dielectric function,
the SP band structure can be also obtained, in perhaps 
a more intuitive way, by writing down a set of coupled 
equations of motion in {\em time} for the multipole moments.  
We first introduce the scalar quantities $q_{n \ell m}$, defined
as the $(\ell m)^{th}$ multipole moment of the $n^{th}$ particle.  Then,
in the absence of damping, the coupled equations of motion can be written
in the form
\begin{equation}
\ddot{q}_{n \ell m} = -\omega^2_{\ell m}q_{n\ell m}
+ \omega_p^2{\sum_{\ell^\prime m^\prime n^\prime}}^\prime 
Q_{n \ell m; n^\prime \ell^\prime m^\prime}q_{n^\prime\ell^\prime m^\prime}
\label{eq:eom}
\end{equation}
where the prime over the sum indicates a sum only over the
grains $n^\prime \neq n$.  For spherical grains, the single-grain resonant
frequencies are given by
\begin{equation}
\omega_{\ell m}^2 = \frac{\ell}{2\ell+1}\omega_p^2,
\end{equation} 
and the coupling constants $Q_{n\ell m; n^\prime\ell^\prime m^\prime}$
are given by Eq.\ (\ref{eq:qlmn}).  Eqs.\ (\ref{eq:eom}) are readily solved
for the eigenfrequencies by substituting assumed solutions of the form
\begin{equation}
q_{n \ell m}(t) = $Re$[q_{\ell m}\exp(inkd - i\omega_\alpha t)]
\label{eq:soln}
\end{equation}
into Eqs.\ (\ref{eq:eom}).  
Here $q_{\ell m}$ is the amplitude of the $(\ell m)^{th}$ multipole in a propagating mode 
of wave vector $k$.
With this substitition, Eqs.\ (\ref{eq:eom}) reduces to a set of coupled homogeneous linear 
algebraic equations. A solution is obtained if the determinant of the matrix of coefficients
vanishes. This condition is equivalent to that of Eqs. (\ref{eq:det}), (\ref{eq:matrix}), 
and (\ref{eq:drude}).

Eq.\ (\ref{eq:eom}) has a straightforward physical interpretation.  
The right-hand side of Eq.\ (\ref{eq:eom}) describes two contributions to
the force acting on the multipole moment $q_{n \ell m}$.  
The first term is the restoring
force due to charge motion within a single particle.
The second term on the right 
comes from the electric fields of all the multipole
fields from the other particles, evaluated at the position of the 
$n^{th}$ particle.   Damping is easily included in Eq.\ (\ref{eq:eom}) by adding to the
right-hand side the term $-\dot{q}_{n\ell m}/\tau$.
To obtain solutions in the presence of damping, we assume 
the form (\ref{eq:soln}) but with a {\em complex} frequency 
$\omega_\alpha(k) = \omega_1 + i\omega_2$.  The resulting $\omega_\alpha(k)$
is given by Eq.\ (\ref{eq:solution}).

An appealing feature of Eqs.\ (\ref{eq:eom}) is that one can easily
add a driving term.  For example, if a uniform electric field 
${\bf E}_n(t)$ is applied to 
the $n^{th}$ grain, the interaction energy between that field and 
the $n^{th}$ grain would be
$H^\prime = -{\bf p}_n\cdot{\bf E}_n(t)$,
where ${\bf p}_n$ is the dipole moment of the $n^{th}$
grain.  To calculate the force on the $n^{th}$ grain, we write
${\bf p}_n = q{\bf x}_n$,
where $q$ is the total electronic charge in the $n^{th}$ grain, 
and ${\bf x}_n$
is its displacement from its equilibrium position.  
The interaction energy between this charge and
the applied field is thus $-q{\bf x}_n\cdot{\bf E}_n(t)$.
The corresponding force on the charge is just 
$q{\bf E} = M\ddot{\bf x}_n$,
where $M$ is the total mass of the electronic charge in the grain.
Thus 
$\ddot{\bf p}_n = q\ddot{\bf x}_n = \frac{q^2}{M}{\bf E} 
= (4\pi a^3 n_e e^2/3m_e) {\bf E} = 
(a^3\omega_p^2/3){\bf E}$,
where $m_e = 3M/(4\pi a^3 n_e)$ is the electron mass, 
$e = 3q/(4\pi a^3 n_e)$ is the magnitude of the electron charge,
$n_e$ is the electron density, and $\omega_p^2 = 4\pi n_e e^2/m_e$
is the squared plasma frequency.

Finally, to incorporate the damping and driving terms into the
right-hand side of Eq.\ (\ref{eq:eom}), we express the applied
electric field in terms of the spherical components $m = 0$ and
$m = \pm 1$.  Thus, we write $E_{n,1,0}(t) = E_{n,z}(t)$,
$E_{n,1, \pm 1}(t) = E_{n,x}(t) \pm E_{n,y}(t)$.  We then obtain the
following equations of motion:
\begin{equation}
\ddot{q}_{n\ell m} = -\omega^2_{\ell m}q_{n\ell m} - \frac{1}{\tau}
\dot{q}_{n\ell m } + \frac{\omega_p^2 a^3}{3}E_{n,1, m}(t)\delta_{\ell,1}  
+ \omega_p^2{\sum_{n^\prime \ell^\prime m^\prime}}^\prime
Q_{n\ell m ; n^\prime\ell^\prime m^\prime}
q_{n^\prime\ell^\prime m^\prime}.
\label{eq:eom1}
\end{equation} 

Eqs.\ (\ref{eq:eom1}) are generalizations of the equations written 
down in Ref.\ \cite{brong} which include all multipole moments, within
the quasistatic approximation, and single-grain damping within
the Drude approximation.  They also include the effects of a
uniform but time-dependent electric field applied to the
$n^{th}$ nanoparticle.

Finally, we note that we have not included radiative corrections to the calculated SP 
bands. However, the present work could also be extended beyond the
quasistatic approximation to include radiative corrections, 
in a simple manner, by adding an additional imaginary part to the 
eigenvalues.  These corrections could easily be included within the dipole 
approximation, as has been discussed, for example, in Ref.\ \cite{brong}.   

\newpage
\begin{figure}
\epsfxsize=10cm \epsfysize=15cm \epsfbox{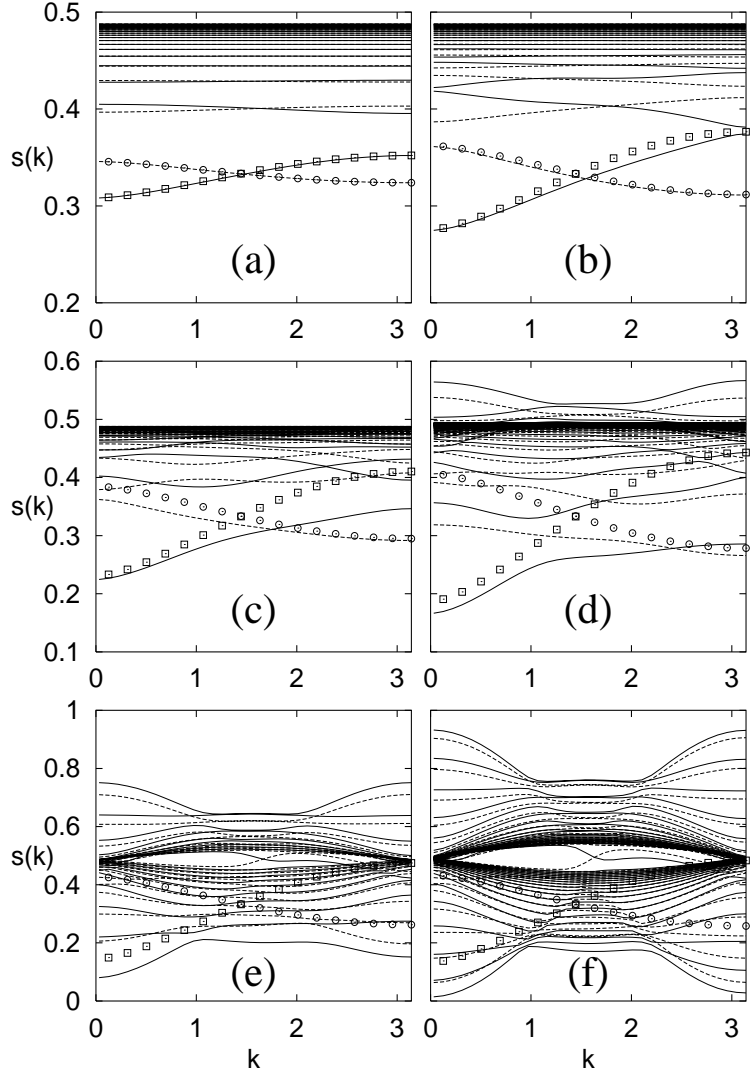}

\caption{Dispersion relations $s(k)$ for the surface
plasmon bands propagating along a chain of spherical nanoparticles
of dielectric function $\epsilon_m$ in a host of dielectric function
$\epsilon_h$, plotted versus wave vector $k$.  (a) $a/d$ = 0.25,
(b) $a/d$ = 0.33, (c) $a/d$ = 0.4, (d) $a/d$ = 0.45, (e) $a/d$ = 0.49,
(f) $a/d$ = 0.5.  The solid and dashed curves correspond to $m = 0$ and 
$m = \pm 1$ respectively for the full band structure, incorporating all
bands up to $\ell = 80$, as obtained diagonalizing the full Hamiltonian
matrix [Eq. (\ref{eq:matrix})].  The open squares ($m= 0$) and circles 
($m=\pm 1$) denote calculations for the $\ell = 1$ modes in the dipole 
approximation.  Note that the $m = \pm 1$ modes are degenerate.}  

\end{figure}

\begin{figure}
\epsfxsize=10cm \epsfysize=7cm  \epsfbox{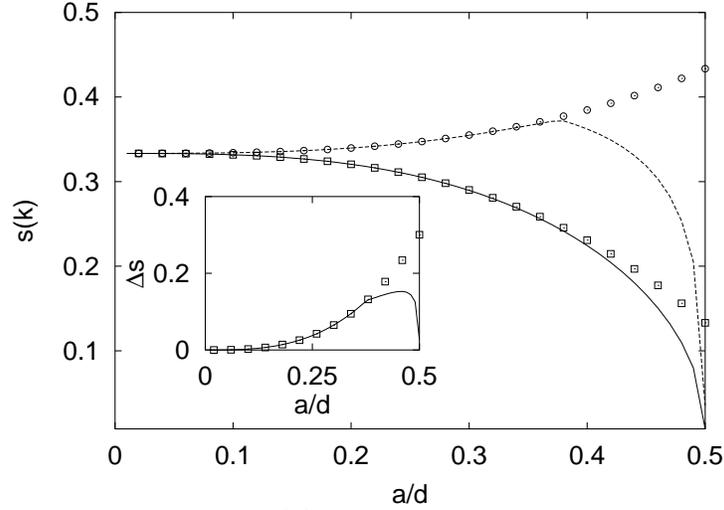}
\caption{Eigenvalues $s(k)$ for the lowest two bands of the band structure
shown in Fig.\ 1 at $k = 0$, evaluated as a function of $a/d$.  
Solid and dashed curves correspond to $m = 0$ and $m = \pm 1$ respectively 
for full multipole calculations.
Open squares ($m=0$) and circles ($m=\pm 1$) are point-dipole calculations.
Inset: Splitting $\Delta s$ between lowest $m = \pm 1$ and $m = 0$ bands at
$k = 0$ as calculated in the dipole approximation (open squares) and using
full Hamiltonian matrix (solid line).}
\end{figure}  

\begin{figure}
\epsfxsize=10cm  \epsfysize=12cm \epsfbox{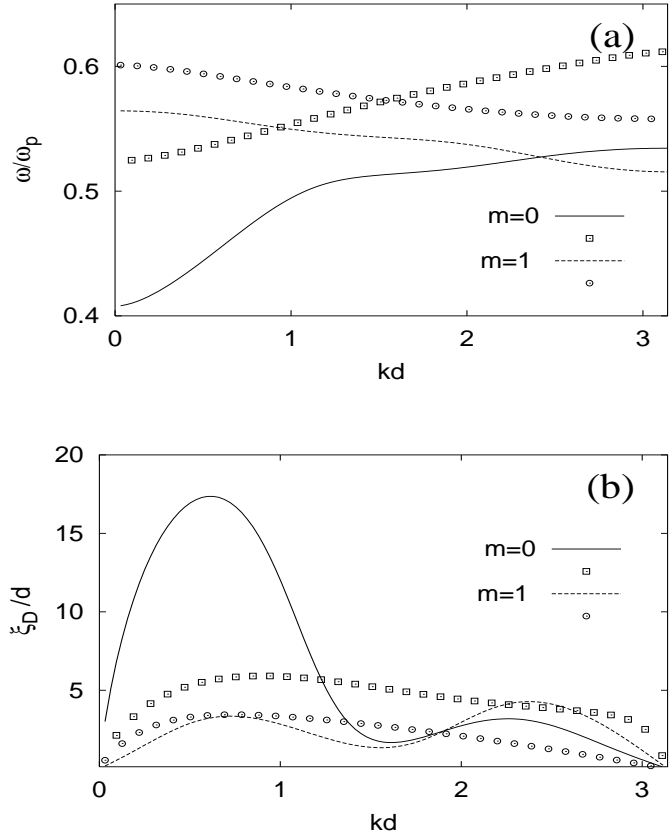}
\caption{(a) Dispersion relations $\omega(k)$ for the lowest two bands
in a chain of metallic nanoparticles at $a/d = 0.33$  or 0.45.  
The solid ($m=0$) and dashed ($m=\pm 1$) lines correspond to $a/d = 0.45$; 
the open squares ($m=0$) and circles ($m=\pm 1$), to $a/d = 0.33$. 
The curves are computed using the full Hamiltonian up to $\ell = 80$, 
using a Drude dielectric function for the metal.
(b) Energy decay lengths $\xi_D$, in units of the lattice constant
$d$, and corresponding group velocities $v_g$ in units of $\omega_pd$,
plotted versus $kd$ for the lowest two bands, assuming
$a/d = 0.33$ or $0.45$.  The labeling of the curves follows the notation 
of Figs.\ 3 (a).}
\end{figure}

\end{document}